\documentclass[
    ,final            
  ]
  {aipproc}
\usepackage{subfigure}
\usepackage{lineno}
\usepackage{url}
\layoutstyle{6x9}

\begin{document}

\title{ATLAS Results from the first Pb-Pb Collisions}

\classification{Pacs:25.75.-q,25.75.Ag,25.75.Dw,25.75.Gz,25.75.Ld}
\keywords      {Relativistic Heavy Ion Collisions, ATLAS experiment, LHC}

\author{Sebastian N. White, for the ATLAS Collaboration}{
  address={The Rockefeller University, NY, USA\footnote{also at INFN Sezione di Pisa and Dipartimento di Fisica E. Fermi, Universit\`a  di Pisa, Pisa, Italy}}
}

\begin{abstract}
 	The ATLAS detector is capable of resolving the highest energy pp collisions at luminosities 
	sufficient to yield 10's of simultaneous interactions within a bunch collision lasting <0.5 nsec. 
	Already in 2011 a mean occupancy  of  20 is often found in pp running. In 2004 studies by ATLAS
	showed that the detector would have excellent performance also for the foreseeable particle multiplicities
	in the highest energy p-Pb and Pb-Pb collisions that the LHC will produce. These studies resulted in a 
	letter of intent to the LHC committee by ATLAS  to do physics with these beams also. In the past 2 years of 
	data taking, ATLAS detector performance studies have confirmed these expectations at the actual multiplicities
	presented below.
	
		The ATLAS program removes an artificial specialization that arose about 30 years ago in high energy
	physics when the energy and intensity frontier moved to colliders. Before that time, for example, the same experiment
	that discovered the $\Upsilon$ (CFS and E605 at Fermilab) also measured the nuclear modification factor in
	the production of high $p_T$ identified charged hadrons using nuclear targets from Beryllium through Tungsten.
 \end{abstract}

\maketitle


\section{Introduction}
	In roughly 1974 interest began to develop in studying the highest energy hadron-nucleus and nucleus-nucleus
	collisions. That year is sometimes identified with a triumph of $e^+e^-$ colliders- the "November Revolution" \cite{November}
	as well as the first papers on QCD\cite{Gell-Mann}. Also in the same year a paper by Lee and Wick asked "whether
it is experimentally possible in a limited domain in space to excite the ordinary vacuum to an abnormal state"\cite{LeeWick}. A conference
	in Bear Mountain, NY, organized by Lee, Lederman, Ruderman et al.\cite{Baym} explored the experimental possibilities,
	including using an existing collider (ISR) with nuclear beams. In the same year interest  developed in hadron-nucleus collisions as
	a tool to understand the space-time development of particle production due to the Fermilab experiments of Busza\cite{Busza} et al.
Since the LHC provides access to both the highest energy pp collisions and the highest energy collisions of nuclear beams it
is natural to use the largest and most complex particle detectors (e.g. ATLAS) to explore these topics also.
	
	ATLAS can also bring about a convergence with the proton and nuclear structure studies traditionally done in electron-nucleus (or proton)
	collisions (i.e. HERA). Both the p-Pb data\cite{Salgado} and the hard photoproduction data\cite{Ramona} will provide unprecedented access to
	low-x structure of hadrons.
\section{ATLAS Detector and Dataset}
\begin{figure}
  \includegraphics[height=.3\textheight]{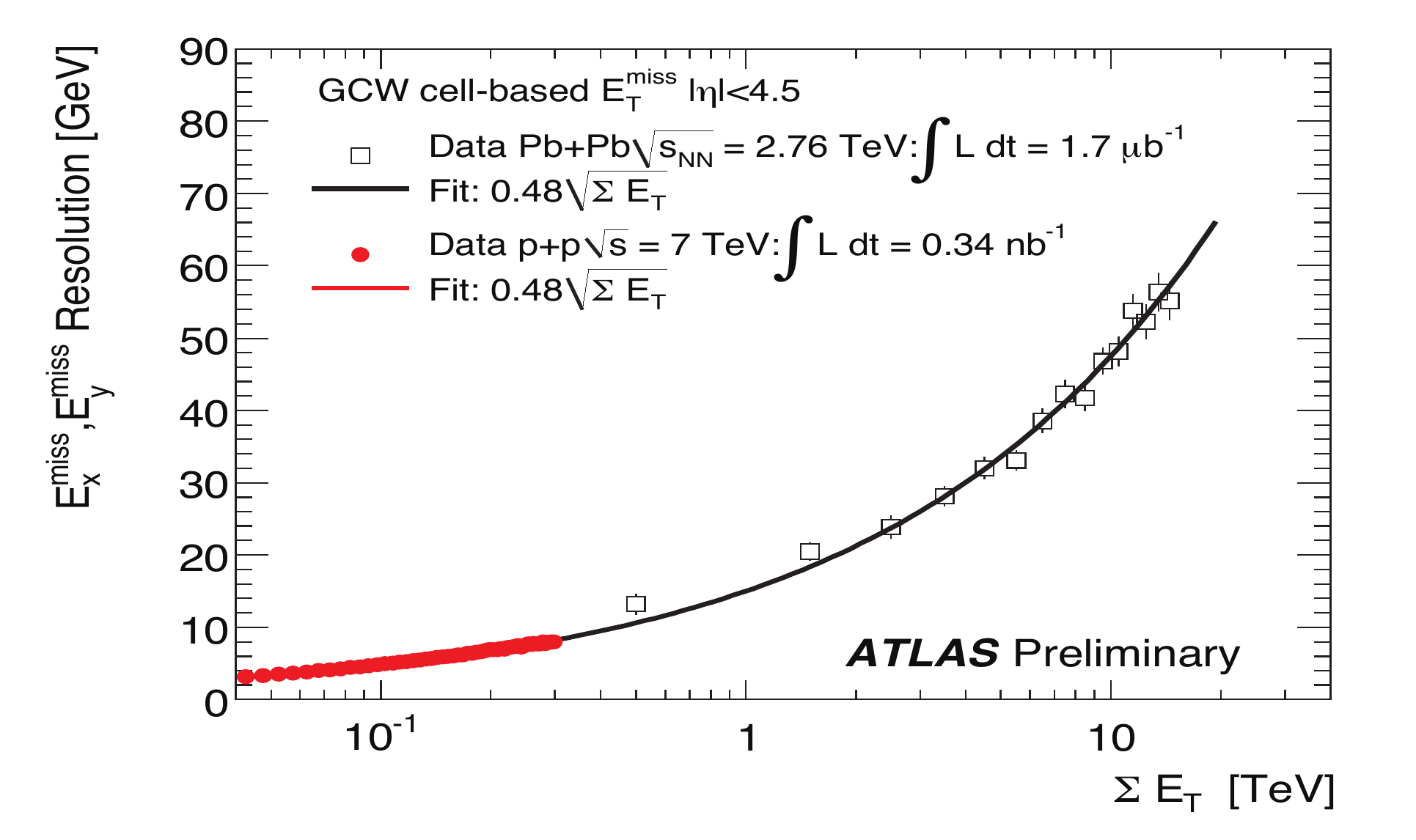}
  \caption{The x and y components of missing $E_T$ resolution vs. $E_T$ showing that the same behavior holds for Pb-Pb data as in pp.}
\end{figure}

	The primary focus in preparing ATLAS for Pb running was to ensure good track reconstruction efficiency and small fake fraction for charged tracks in the 
	inner detector for the highest multiplicity events. This dominates the track quality for measurement of charged multiplicity, nuclear modification of the 
	charged particle $p_T$ spectra as well as for the reconstruction of muons (since the outer tracking region for muons always has low occupancy, as expected). The inner tracking efficiency
	was well modeled by simulations resulting in an uncertainty of $3\%$.
	
		The ATLAS calorimeter is highly segmented, both longitudinally and transversely and covers the full rapidity region to $\pm 4.9$. Nevertheless in the
	highest multiplicity Pb-Pb events the underlying event, due the high multiplicity of soft particles, will typically contribute to the energy deposit in a given 
	tower being used to measure jet energy, for example. The underlying event subtraction must be determined from the data themselves, as discussed below. It is important
	that the calorimeter response vs. energy deposit be well understood. This is demonstrated in Fig. 1, where the x and y components of the missing transverse 
	energy resolution are plotted for both pp collisions (which have lower transverse energy than is possible in Pb-Pb) and for Pb-Pb collisions.
	
	In the 2010 Pb-Pb run ATLAS recorded approximately 9 $\mu b^{-1}$ of luminosity at a recording efficiency of $\sim 95\%$. Of these data, roughly $10\%$ were
	recorded with central magnetic field turned off in order to increase the efficiency of low momentum tracking for the charged multiplicity measurement , discussed below. The minimum bias trigger consisted of the
	Minimum Bias Trigger Scintillators ( MBTS, covering the region $2.09\leq|\eta|\leq3.84$) since these were familiar from the pp runs and, for most of the data, Zero Degree Calorimeters
	which subtend an angle of $\sim350$ $ \mu$rad with respect to the beam direction and accept beam energy neutrons from nuclear breakup out to $p_T\sim 1 $ GeV/c.
	
	All data were collected with a beam energy of 1.38 TeV/nucleon and a typical luminosity of a few$\times 10^{25}$ yielding an inelastic collision rate of 100's of Hz.
\section{Measurement of Charged Particle Multiplicity and correlations}

\begin{figure}
  \includegraphics[height=.5\textheight]{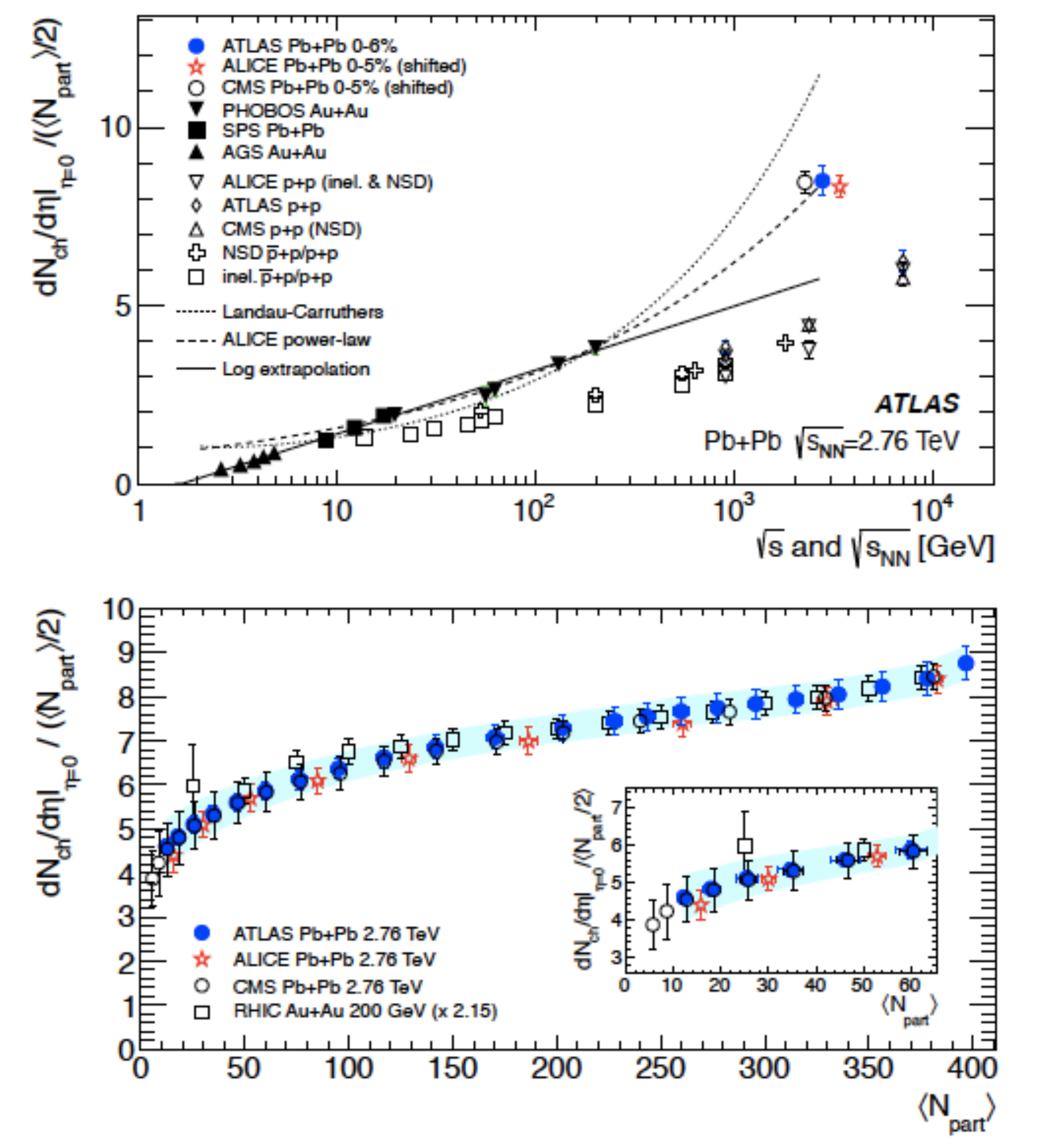}
 
  \caption{Charged multiplicity density per participant nucleon pair for the most central ATLAS collisions (top). Dependence on the number of participants (bottom).}
\end{figure}

	Charged particle multiplicity and correlations are more difficult to interpret in pp data than in Pb-Pb because, up to now, the experimental tools for characterizing the initial 
	conditions of pp collisions have been limited. In contrast, for the Pb-Pb data presented here the initial collision impact parameter (or, by calculation, the number of participant nucleons ) and orientation of the collision are determined with good precision. 
	
	In the case of pp data, for example, multiplicity is measured for a sample of collisions ranging from peripheral (and, in some cases, diffractive) to very central for which we would expect more or less copious particle production. Similarly the leading baryon effect implies that a large fraction of the collision energy is carried away by a leading baryon (protons or neutrons with roughly equal probability) in each hemisphere. It has been shown that when the effective c.m.s. energy of pp collisions is corrected for the leading particle effect, pp multiplicities exhibit the same
trend vs. energy as those from $e^+e^-$ or even nucleus-nucleus\cite{Carruthers}. 
	
	In the case of Pb-Pb collisions the impact parameter is determined by mapping the frequency distribution in any sensitive variable into the known frequency distribution of collisions vs. impact
	parameter. ie:
\begin{equation}
Luminosity(b)\propto b\times db
\end{equation}
	In the present analysis the frequency observed in $\Sigma E_T$ in the forward calorimeter is used. The charged multiplicity is then renormalized by the calculated number of participant nucleons for this collision\cite{Multiplicity}.

\begin{figure}
  \includegraphics[height=.5\textheight]{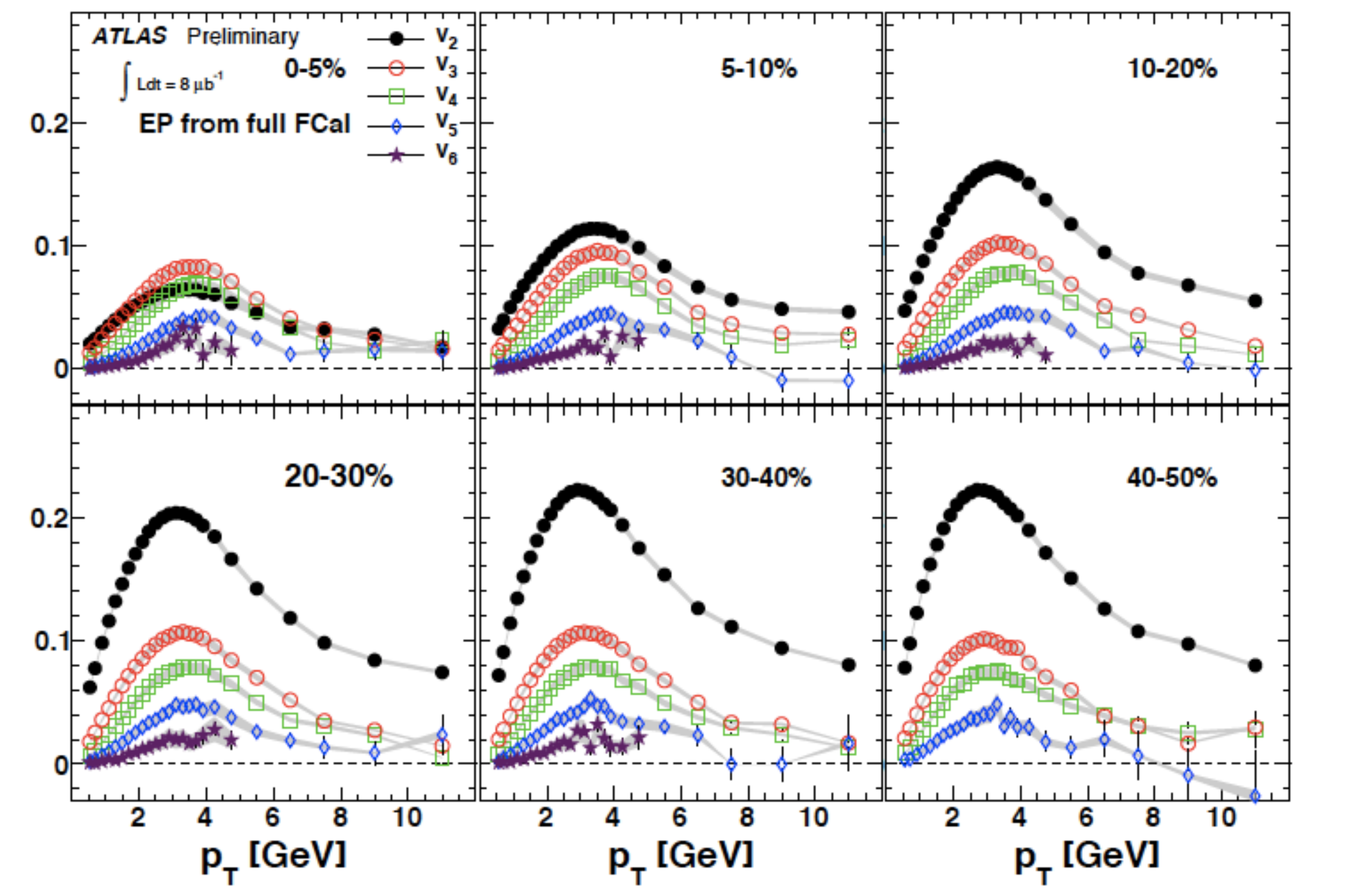}
 
  \caption{$v_n$ vs pT for several centrality classes. }
\end{figure}

\section{Correlations}

	In Pb-Pb collisions the azimuthal distribution in either the forward or backward (or both) hemispheres
can be used to learn about the initial geometry of the collision. Because the reaction plane (i.e. the
orientation of the impact parameter vector, b) is independently measured in each  hemisphere, the
resolution of the reaction plane can be determined directly from our data. One method to measure the
reaction plane (full FCAL method) consists of determining the plane in each forward calorimeter (3.2$\leq|\eta|\leq 4.9$)
from the energy distribution in the calorimeter cells of the first 2 depth layers and combining the 2 measurements.
	The azimuthal distribution of charged particles measured in the inner tracker can be expressed as
a Fourier series, ie
\begin{equation}
\frac{dN}{d\Phi}\propto\Sigma v_n Cos(n\Phi)
\end{equation}

where $\Phi$ is the azimuthal angle of tracks relative to the reaction plane. The first moment of the distribution,
called directed flow must by symmetry, equal zero at central rapidities so it is not presented in the analysis here
from the inner tracker. We find that the flow moments vary little with pseudorapidity but depend strongly on track $p_T$
and on centrality (given by equation 1). The panels in Fig. 3 are divided into centrality classes with the upper left taken
from the 5$\%$ smallest impact parameter events\cite{Correl}.
	
		The flow moments, $v_n$, are sensitive to the hydrodynamic properties of the hot, dense precursor to hadron
production in these events. One hydrodynamic calculation\cite{Shuryak} attempts to account for the relative strength of the $v_n$
for various choices of viscosity of this medium and obtains approximate agreement.

\section{Jet Quenching}

\begin{figure}[ht]
\begin{minipage}[b]{0.5\linewidth}
\centering
\includegraphics[scale=.47]{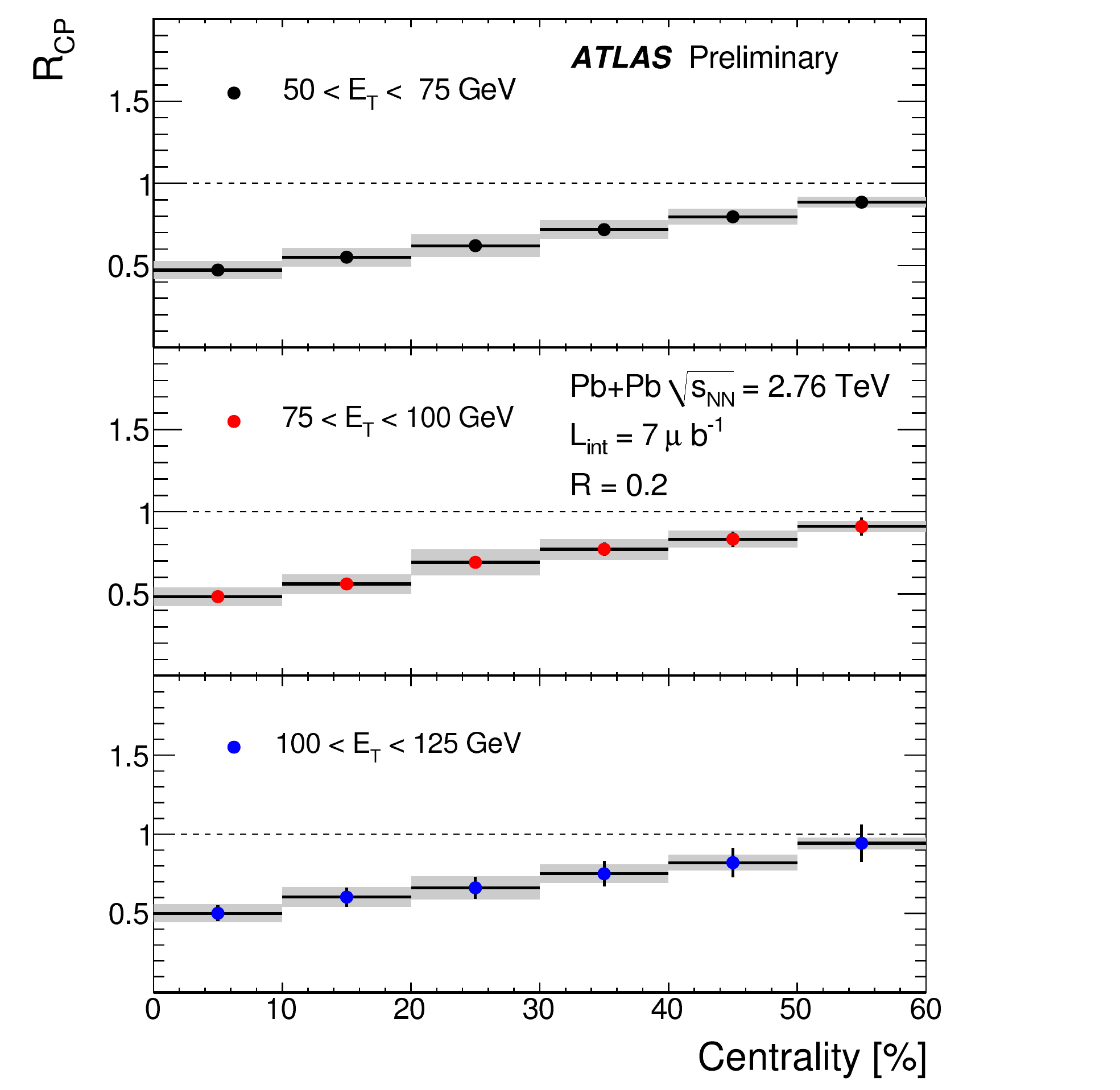}
\caption{The jet yield for various centrality classes divided by yield in a peripheral class(=$R_{CP}). R=02, 0.4.}
\label{fig:figure1}
\end{minipage}
\hspace{0.5cm}
\begin{minipage}[b]{0.5\linewidth}
\centering
\includegraphics[scale=.47]{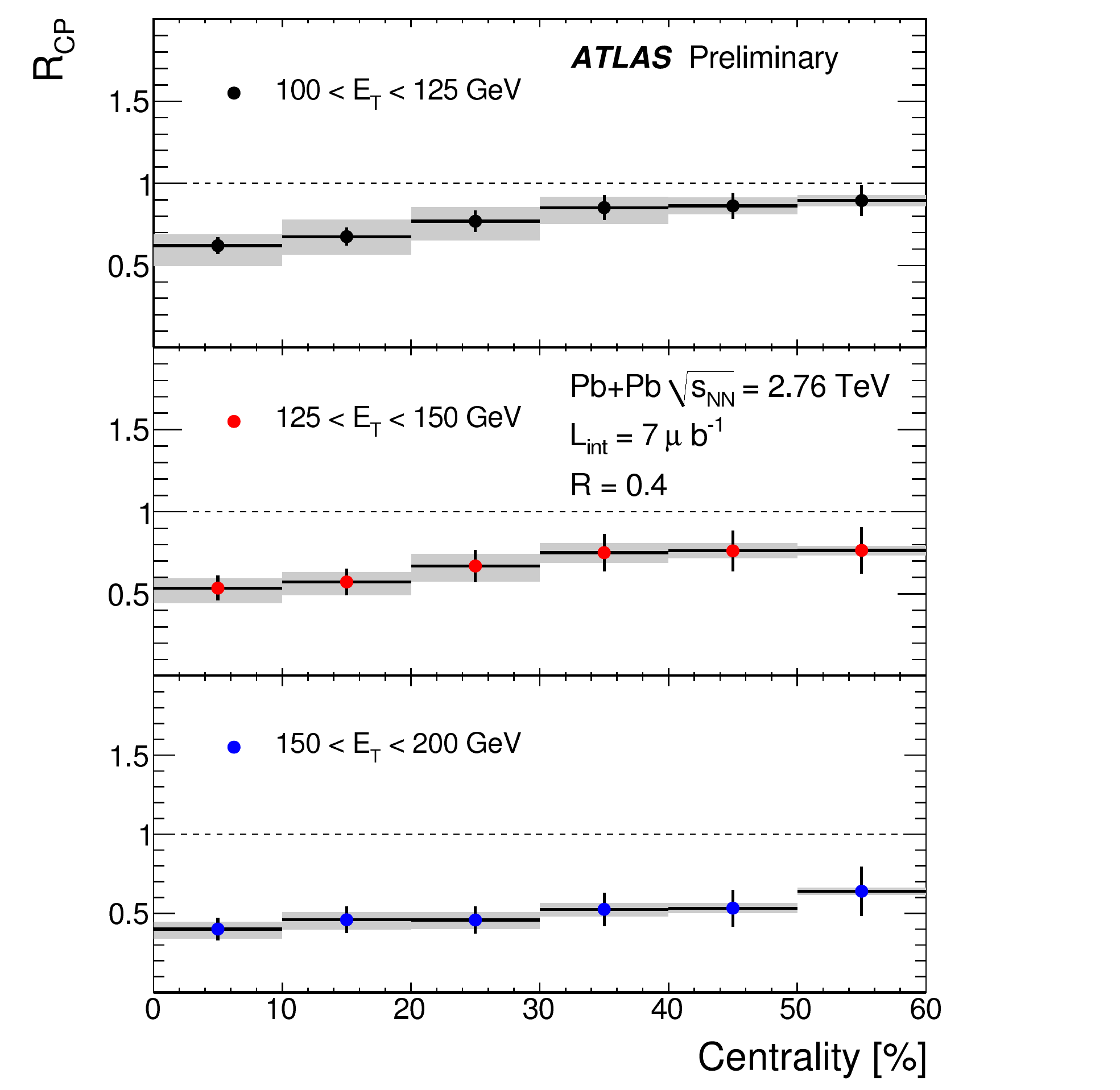}
\caption{jet yield for various centrality classes divided by yield in a peripheral class. R=02, 0.4.}
\label{fig:figure2}
\end{minipage}
\end{figure}

	Hard scattering of partons within Pb-Pb collisions can be used to probe the field of high parton density within which the scattering
occurs since the hard scattering occurs at the very beginning and the scattered parton must traverse the medium before it emerges
and fragments into a jet. There are a number of ways that the parton could lose energy as it traverses the medium, including radiative
and collisional energy loss. Although fragmentation occurs late in time compared to the time it takes the parton to traverse this hot
medium,  gluons radiated during the energy loss process would effectively modify the particle distribution (fragmentation functions) of
particles in the reconstructed jet.

	If the energy loss is purely the result of elastic collisions the jet fragmentation could look like that of a jet produced in a pp collision
but the reconstructed jet axis might be deflected. This would lead to a modified distribution in the $\Phi_{Jet-Jet}$ correlation of back-to-back jets.
In fact, ATLAS sees small modification of the back-to-back correlation of di-jets in central collisions (shown elsewhere\cite{ATLASjets}).
Studies of jet modification in central Pb-Pb collisions could shed light on the process of vacuum fragmentation, which is essentially treated
as an empirical parametrization in most of high energy physics.

	Jet reconstruction in heavy ion collisions is more difficult in Pb-Pb data than in pp and doing it correctly has been a focus ever since the 
original studies for the ATLAS Heavy Ion LOI. As has been pointed out in the literature, incorrect treatment of the underlying event energy and
its fluctuations can easily lead to artifacts\cite{Salam}.

	The method used by ATLAS to reconstruct jets in this environment attempts to do this making the best use of the detailed measurement of 
energy flow in ATLAS'  large acceptance and highly segmented calorimeter.

 \begin{figure}[ht]
\begin{minipage}[b]{0.5\linewidth}
\centering
\includegraphics[scale=.43]{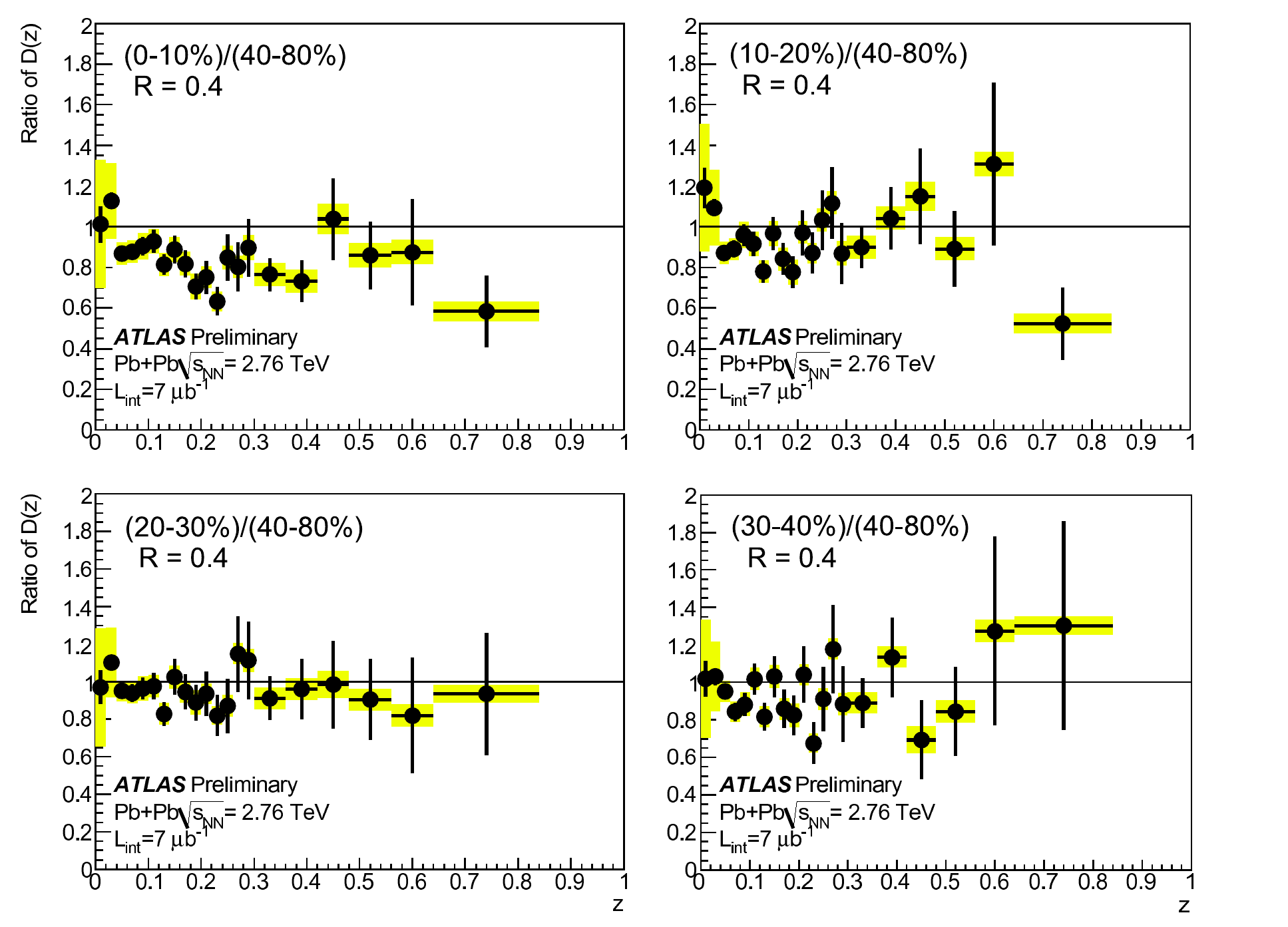}
\caption{Longitudinal (left) and transverse fragmentation distributions for various centrality classes divided by a peripheral class. R=0.4}
\label{fig:figure1}
\end{minipage}
\hspace{0.5cm}
\begin{minipage}[b]{0.5\linewidth}
\centering
\includegraphics[scale=.43]{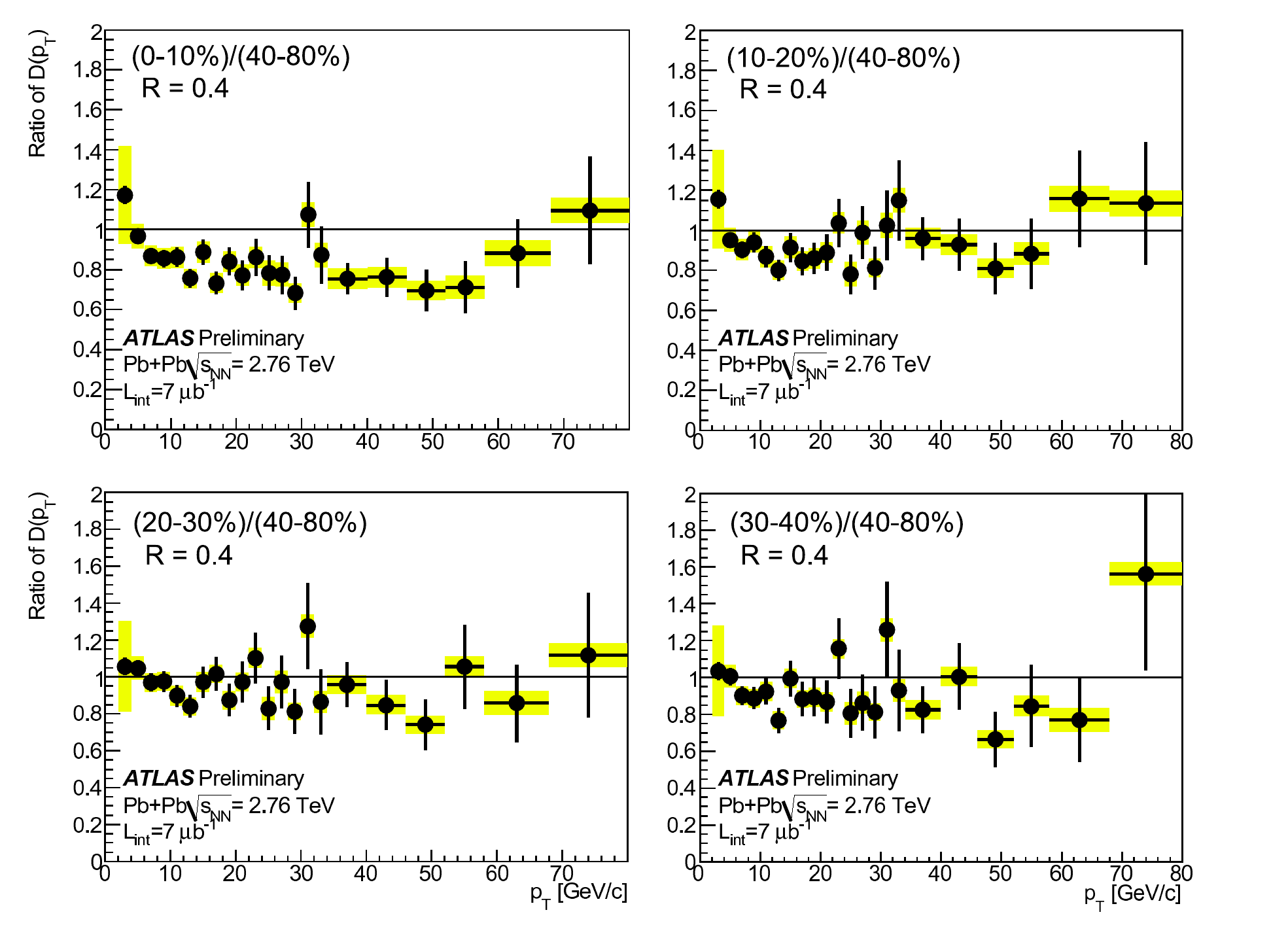}
\caption{Longitudinal (left) and transverse fragmentation distributions for various centrality classes divided by a peripheral class. R=0.4}
\label{fig:figure2}
\end{minipage}
\end{figure}

	For results presented here jets are reconstructed using an anti-$k_t$ algorithm and cone radii of both R=0.4 and R=0.2. Layers of the calorimeter
within a tower of $0.1\times0.1$ are combined taking into account energy-density dependent corrections arising from non-compensation of the calorimeter.  We
then calculate a distribution of energy flow from the underlying event, based on energy deposit excluding the jet window. The modulation, due to flow 
of particles in the underlying event is determined using the results presented above and the reaction plane determined with the full FCAL method.
This method, nevertheless, slightly overcompensates since a fraction of the jet energy contributes to the normalization of the underlying event energy.
A second iteration of the subtraction procedure is therefore performed. Results presented here are for jets with $|\eta_{jet}|\leq 2.8$.

	Fluctuations of the underlying event (tower-by-tower and for different layers) are obtained from simulation with HIJING+GEANT and compared to 
data. Our conclusion is that fluctuations in the data are reproduced to within $10\%$.

	Complementing an earlier publication which showed evidence of jet quenching using di-Jet events we have measured inclusive jet production
and compared the yield in several centrality bins to the yield in peripheral ($80\%$) collisions. Figure 4 shows the yield expressed as a ratio of central
to peripheral. The suppression reaches $\sim 0.5$ independent of jet energy for both R=0.2 and 0.4 jet cone radii. This is a similar suppression to what was 
observed for high $p_T$ charged particles at RHIC, which leads one to suspect that the fragmentation function for jets produced in central Pb-Pb
collisions is unmodified from the vacuum distribution.

	Fig. 5 shows that this is true in detail for the case of R=0.4 reconstructed jets. Similar results are obtained for R=0.2\cite{ATLASjets}.	
	
\section{J/$\psi$ Suppression and $W^{\pm}$ reference measurement}

	Since the jet yield is modified so dramatically it would not be surprising to see suppression of a di-quark bound state, such as the J/$\psi$  .
In fact this has been observed early in the SPS Heavy ion program by NA50. There are many aspects of the observed suppression which are not 
well understood, however. Originally it was felt that this suppression must occur because color screening in Heavy Ion collisions would prevent
formation of a quark-antiquark bound state. One way to get further insight into this problem is to search for similar suppression of $Z^0$ or $W^{\pm}$ bosons
since these are neither strongly interacting or composite. Any modification of their yield would have to arise from the initial state (i.e. due to nuclear structure function).

	ATLAS has both measured J/$\psi$ suppression and searched for similar suppression of the weak vector bosons.
	
	Dimuons were selected with opposite charge,  $p_T\geq$3 and $|\eta|\leq$2.5. After sideband subtraction to
remove dimuons from the continuum a sample containing 190$\pm$ 20 J/$\psi$ candidates was obtained for the most central $10\%$.
A suppression in this most central bin relative to a peripheral sample of $\sim 0.5$, is found which is not very different from the jet suppression result.
	
	To obtain a sample of $W^{\pm}$ candidates muons were selected by removing events with a high mass dimuon ($m_{\mu\mu}\geq 66$ Gev) and the
inclusive spectrum was fit to combination of the continuum spectrum -mostly due to charm and b quark decays- and a W-decay template tuned on
a W sample from pp data. The inclusive spectrum was then fitted to obtain the number of W candidates in a given centrality class of Pb-Pb data,
as illustrated in Fig. 6. Also shown in Fig. 6 is the yield ratio for peripheral collisions divided by the yield in a given centrality class, which shows
consistency with what you would expect from binary collisions scaling- i.e. no $W^{\pm}$ suppression.

 \begin{figure}[ht]
\begin{minipage}[b]{0.5\linewidth}
\centering
\includegraphics[scale=.43]{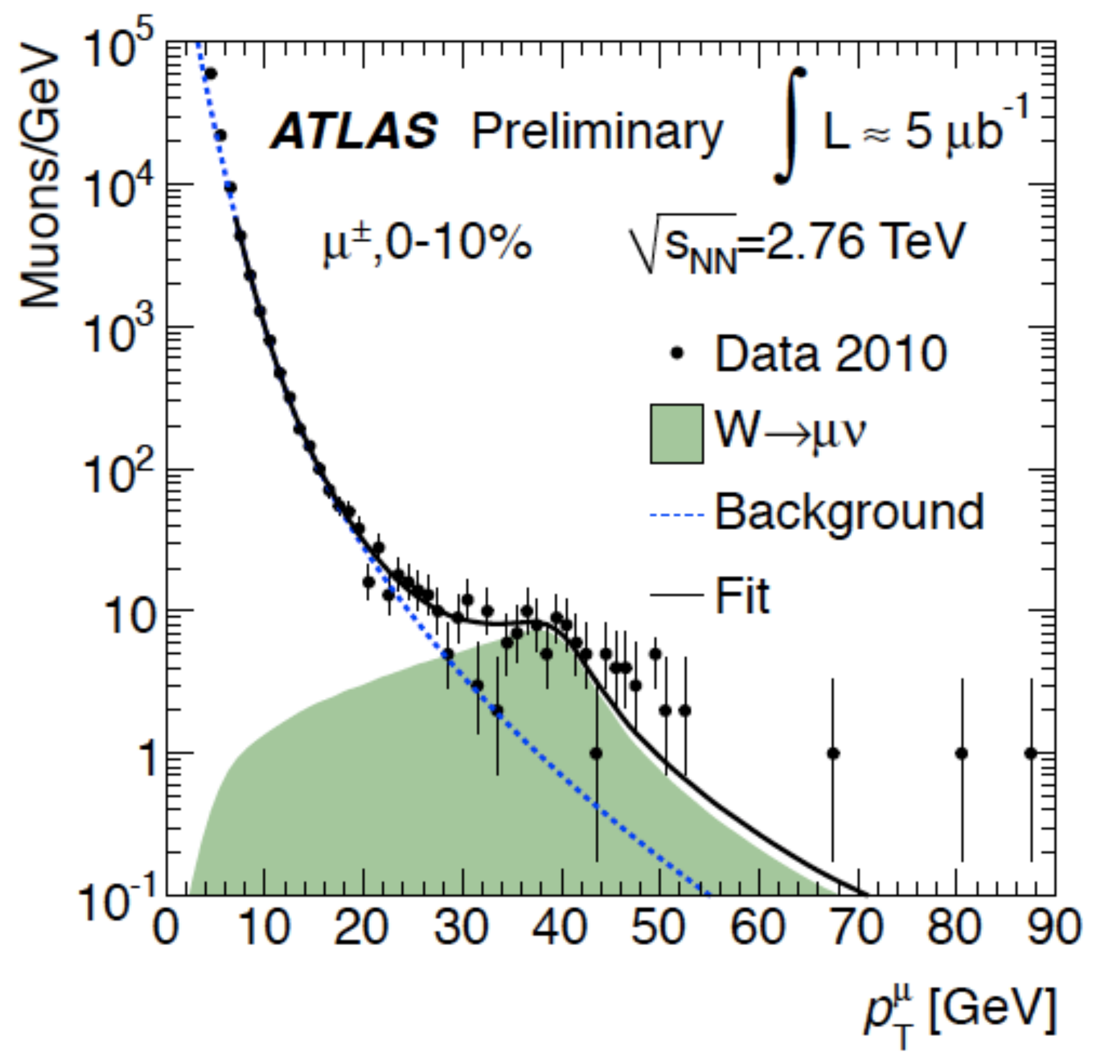}
\caption{ (left)Inclusive muon $p_T$ distribution with cuts described in the text. Also shown is the fit to $W$ decay template and the background from heavy flavor decays.
(right) The $W$ yield vs. centrality selection presented as peripheral yield divided by yield in the centrality bin, corrected for the number of binary collisions. }
\label{fig:figure1}
\end{minipage}
\hspace{0.5cm}
\begin{minipage}[b]{0.5\linewidth}
\centering
\includegraphics[scale=.43]{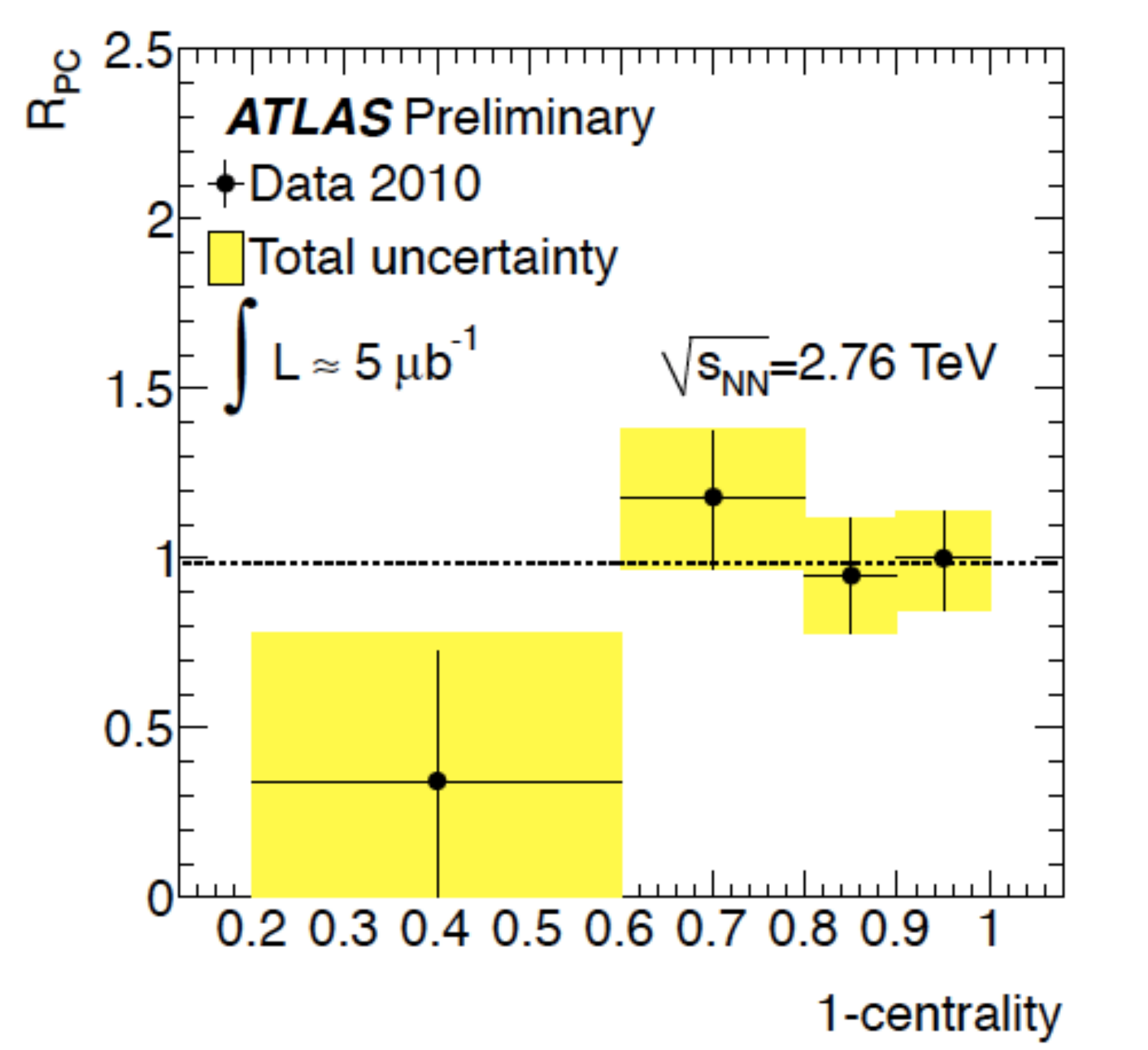}
\caption{ (left)Inclusive muon $p_T$ distribution with cuts described in the text. Also shown is the fit to W decay template and the background from heavy flavor decays.
(right) The $W$ yield vs. centrality selection presented as peripheral yield divided by yield in the centrality bin, corrected for the number of binary collisions. }
\label{fig:figure2}
\end{minipage}
\end{figure}

\section{Outlook}

	In November 2011 the LHC will again collide Pb beams  (yielding an expected $\sim4$-fold increase in integrated luminosity) and for a short test, possibly yielding of order $10^6$ inelastic collisions, will attempt p-Pb collisions also.
		
	There are a number of further measurements that could be used to clarify the mechanism of jet quenching. CMS has shown that the energy loss
that results in jet quenching appears in the form of low $p_T$ particles at wide angles to the jet. The anisotropic emission of particles described by the $v_n$ moments is 
modeled as small initial state perturbations which propagate hydrodynamically. It would be interesting to study the $v_n$ for a class of events where a $\sim$ 50 GeV jet is emitted 
since its energy loss must produce a large perturbation. It would also be of interest to compare jet suppression for b-quark jets.

	Near $\eta$=0, d-Au data at RHIC showed an enhancement rather than a suppression of high $p_T$ charged particles as did E605\cite{E605}. However at forward rapidities a
suppression was observed. New data on p-Pb will be useful for discriminating among present models for this suppression. CMS has shown that similar features of
correlations in Pb-Pb data also appear in pp collisions selected for high multiplicity which likely corresponds to very large gluon densities\cite{CMS} and studies are
under way which reveal other similarities to high multiplicity Pb-Pb collisions\cite{Dremin}. It would be interesting to explore other methods\cite{Otranto} to select pp collisions 
with similar properties.

\begin{theacknowledgments}

 We thank the LHC operations group for the excellent performance  of the machine during
 this first run with Pb beams. I would like to thank the organizers of the EPIC conference for this excellent
 opportunity to discuss the interpretation of these new results and the organizers of the Galileo Galilei Institute workshop
 on "QCD after the start  of the LHC" for their hospitality during the preparation of this manuscript. Copyright CERN for the benefit of the ATLAS Collaboration.

 \end{theacknowledgments}

\bibliographystyle{aipproc}   

\end{document}